\documentclass[letterpaper]{article}

\usepackage{aaai}
\usepackage{times}
\usepackage{helvet}
\usepackage{courier}
\usepackage{amssymb,amsfonts,amsmath,amscd,array,bm}
\usepackage{multirow}
\usepackage{graphicx}
\usepackage{listings}
\usepackage{nicefrac}
\usepackage{enumerate}
\usepackage{setspace}

\usepackage{subfig}
\usepackage{tikz}
\usepackage{pgf}
\usepackage{pgffor}
\usepgflibrary{plothandlers}

\usepackage{natbib}

\def\urltilda{\kern -.15em\lower .7ex\hbox{\~{}}\kern .04em}

\def\muacro#1{\texttt{#1}}           				

\def\eqref#1{Equation~(\ref{eq:#1})}

\def\figlabel#1{\label{fig:#1}}
\def\figref#1{Figure~\ref{fig:#1}}

\def\seclabel#1{\label{sec:#1}}
\def\secref#1{Section~\ref{sec:#1}}

\def\tablabel#1{\label{tab:#1}}
\def\tabref#1{Table~\ref{tab:#1}}



\def\vect#1{{\bm{#1}}}

\def\VSP#1{\vect{\mathcal{#1}}}

\newcommand{\argmax}[1]{\underset{#1}{\operatorname{arg}\,\operatorname{max}}\;}

\begin{document}
%
\title{Political Disaffection: a case study on the Italian Twitter community}
\author{C. Monti$^1$, A. Rozza$^2$, G. Zappella$^3$, M. Zignani$^1$, A. Ardvisson$^4$, M. Poletti$^4$\\
\footnotesize $^{1}$ Dipartimento di Informatica, Universit\`a degli Studi di Milano \\ 
\footnotesize	\{monti, zignani\}@di.unimi.it\\
\footnotesize	$^{2}$ Dipartimento di Scienze Applicate, Universit\`a degli Studi di Napoli ``Parthenope''\\
\footnotesize	alessandro.rozza@uniparthenope.it\\
\footnotesize $^{3}$ Dipartimento di Matematica, Universit\`a degli Studi di Milano \\ 
\footnotesize	giovanni.zappella@unimi.it\\
\footnotesize $^{4}$ Dipartimento di Scienze Politiche, Universit\`a degli Studi di Milano \\ 
\footnotesize	\{adam.ardvisson, monica.poletti\}@unimi.it\\
}
\maketitle

\begin{abstract}
	\begin{quote}
		In our work we analyse the political disaffection or ``the subjective feeling of powerlessness, cynicism, and lack of confidence in the political process, politicians, and democratic institutions, but with no questioning of the political regime'' by exploiting Twitter data through machine learning techniques.
  
In order to validate the quality of the time-series generated by the Twitter data, we highlight the relations of these data with political disaffection as measured by means of public opinion surveys. Moreover, we show that important political news of Italian newspapers are often correlated with the highest peaks of the produced time-series.
	\end{quote}
\end{abstract}

\section{Introduction}\seclabel{Introduction}
Twitter is one of the biggest micro-blogging services in the world. Micro-blogging refers to the publication of short text messages, used to share all kinds of information; on Twitter, these messages are called ``tweets'' (their maximum length is $140$ characters), and many millions of them are posted every day.

Twitter is an interesting data source to explore public sentiment trends (\cite{Bollen2011,Indignados}): its content is easily available, and it has a very flexible nature due to the fact that it is currently used for open conversations, public opinions, and news commentaries. Another crucial characteristic of Twitter content is its timeliness; this peculiarity guarantees that tweets are related to a much closer temporal window with respect to other user-generated texts, such as blogs. 

Modelling trends from Twitter data has became a popular research task. Among such studies, those drawing attention to political topics are some of the most attractive, and in the last years a great deal of research has focused on them.

In this study we want to concentrate on an important concept in political science: \textit{political disaffection}, 
or ``the subjective feeling of powerlessness, cynicism, and lack of confidence in the political process, politicians, and democratic institutions, but with no questioning of the political regime''~\citep{DiPalma1970}. In political science, levels of political disaffection are understood to relate to levels of  political participation and, consequently to have important implications for the legitimacy of democratic political systems. This makes the study of political disaffection one of the key topics of contemporary studies of political behavior.
Despite the popularity of the term in the relevant literature, political disaffection is a complex and multi-dimensional term. The concept of political disaffection does not necessarily imply low levels of satisfaction with the current government, nor adversity to the system of democratic politics. Instead,  concept of political disaffection has two components. One, low trust  in politics or politicians in general. Two, what is known as political inefficacy, "a belief that politics is complex, difficult to understand, self-referential and distant from citizens"~\citep{Campbell1954}. 

To our knowledge political disaffection has never been studied using Twitter data.
In this work we propose an automatic approach to measuring political disaffection using Twitter data with the aim to study the relations between our measurement of political disaffection and political disaffection as measured by public opinion surveys.
In accordance with the literature we define political disaffection as a general discontent with the political system, and not as a partisan position against a particular party, individual or policy, we operationalize this definition in three steps. One, we use a supervised methodology to extract a subset of political tweets form the universe of tweets in italian. Two, we perform a sentiment analysis to analyze political tweets with negative sentiment. Three, we automatically select the tweets that refer to politics of politicians in general, rather than specific political events or personalities. 
In order to validate our operational measurement, the selected tweets that represent political disaffection are used to create time series that are related to indicators of political disaffection in public opinion surveys. 
Furthermore, we also show that important political news from Italian newspapers are often correlated with peaks in the produced time-series.  
 
This paper is organized as follows:
in~\secref{Related} the related works are summarized; in~\secref{dataset} we describe the procedures used to generate the datasets employed to train our supervised methods, the approach to extract the tweets employed in our analysis, and we summarize the public opinion surveys used to validate the quality of our approach;
in~\secref{classification} we describe the employed feature extraction methodologies and the classification approach that we used; in~\secref{results} we present the achieved results; ~\secref{Conclusion} highlights the conclusions of our work.

\section{Related Works}\seclabel{Related}
In literature a great deal of research has focused on the analysis of different phenomena using the data of micro-blogging services. Among them, we recall the work proposed in~\citep{Popescu2011} where the authors explore the correlation between types of user engagement and events about celebrities using Twitter data. Furthermore, ~\citep{Bollen2011b} propose an approach able to predict the stock market by employing micro-blogging data.

The most closely related works are those concerning the analysis of political topics by employing Twitter data. 
In~\citep{Bollen2011}, the authors propose a method able to extract from tweets data different time series corresponding
to the evolution of $6$ emotional attributes (tension, depression, anger, vigour, fatigue, and confusion) called Profile of Mood States (\muacro{POMS}).
The authors applied \muacro{POMS} to suggest that socio-economic agitations caused significant fluctuations of the mood levels.

One of the earliest papers discussing the feasibility of using Twitter data as a substitute for traditional public opinion surveys has been proposed in~\citep{Oconnor2010}.
The authors employ Opinion-Finder\footnote{Opinion-Finder is a system that performs subjectivity analysis, automatically identifying when opinions, sentiments, speculations and other private states are present in text.} to determine both a positive and a negative score for each tweet in their dataset.
Then, raw numbers of positive and negative tweets regarding a given topic are used to compute a sentiment score. Using this method, sentiment time series are prepared for a number of topics such as: presidential approval, consumer confidence, and US $2008$ Presidential elections. 
According to the authors both consumer confidence and presidential approval public opinion surveys show correlation with the Twitter sentiment data
computed with their approach. However, no correlation are been found between electoral public opinion surveys and Twitter sentiment data.

In~\citep{Tumasjan2010} an analysis of the tweets related to different parties running for the German $2009$ Federal election is presented. Moreover the authors show that the count of tweets mentioning a party or a candidate accurately reflected the election results suggesting a possible approach to perform an electoral prediction. Furthermore, in~\citep{Livne2011} a novel method that tries to predict elections has been presented. This approach relies both on Twitter data and also in additional information such as the party a candidate belongs to, or incumbency.  In~\citep{Bermingham2011} is proposed a methodology that extends  the previous approaches to incorporate sentiment analysis to perform a prediction on the political election. The authors test their method in the $2011$ Irish General Election finding that it is not competitive when compared with traditional public opinion surveys. Similar approaches are proposed in~\citep{Tjong2012,Skoric2012}.

Nevertheless, some works present some doubts about the possibility to perform an electoral prediction using Twitter data~\citep{Metaxas2011,Avello2012}.
More precisely, in~\citep{Metaxas2011} the authors analyze the results from a number of different elections and they conclude that Twitter data is only slightly better than chance when predicting elections.

\bigskip
In our work we analyse the well-known political attitude of political disaffection by employing Twitter data through machine learning techniques. 
In order to validate the quality of the information extracted from the Twitter data, we highlight the relations of these data with political disaffection as measured in public opinion surveys. Here, the attitude of political inefficacy is used as proxy of this broad concept.
This attitude expresses the (subjective) sense of powerlessness of citizens in politics. Its symmetrical concept, political efficacy, is instead the individual self-image as an influential participant in politics, ``the feeling that individual political action does have, or can have, an impact upon the political process...the feeling that political and social change is possible, and that the individual citizen can play a part in bringing about this change''~\citep{Campbell1954}. Political efficacy is considered crucial for participation since efficacious citizens have a greater propensity to engage in political action, and high levels of political efficacy within a population are important in creating support for the political system. If increasingly more citizens believe that they have not enough abilities to influence political decision-making and opportunities to participate in politics, whilst simultaneously not believing in the accountability of how the political system works, they will become frustrated and discontent, reducing their likelihood to act politically or to vote. Together with political inefficacy we also use low intentions to vote as a proxy of political disaffection, since we believe it is a behavioural consequence of citizens' sense of powerlessness (see~\cite{Torcal2006}). 

\section{Dataset}\seclabel{dataset}
In this section we describe how we have generated the dataset employed to train our supervised methods. Furthermore, the procedure used to extract the tweets employed to identify political disaffection is described. Finally, the information extract from different Italian newspapers, and the employed public opinion surveys are described. 
\subsection{Training Data}
In order to train the classifiers that compose our methodology to extract the political disaffection, we build the training set employing the Twitter API v$1$ by a $2$-step procedure involving a semi-automatic search method and a labelling phase guided by experts. The collection phase began at the beginning of April 2012 and ended at the beginning of June 2012 and collected about $120,000$ of tweets and retweets. The collected tweets resulted from a geo-localized trending topic\footnote{Trending topics are the most popular and talked-about words and phrases on Twitter for a specif time period.} search and a targeted search on political themes. In particular, at the end of each day we requested the top $10$ trending topics for Italy. As most trending topics regard non-political arguments (i.e. celebrities, sports or viral hashtags) we selected the political ones and a subset of the non-political. Furthermore, in order to have a more meaningful number of political tweets, we searched for tweets related to politicians, political news from online newspapers and talk-shows. As query keywords we choose the name of the most famous Italian politicians, the topics of the top news in the political section of online newspapers and the official hashtags of TV-talks. After the deletion of retweets, the resulting dataset is made up by a large corpus of $40,000$ records consisting of the tweet text, its date and the keyword used in the search. 
\par Once the dataset was collected, we started the labelling phase employing the expertise of a pool of $40$ sociology and political science students. Each student was assigned a set of $3000$ tweets to be classified by means of a web application. Two different labels have been associated to each tweet, the first is political/not political (the students need to identify if the tweet topic is political), and the second is positive or neutral sentiment versus negative sentiment of the political tweets. As the meaning of the labels are quite fuzzy, before the labelling process we made a kick-off meeting to ensure a label definition agreement and consequently an acquisition of homogeneous and reliable data. 
\par The tweets' assignment has been made so that each tweet text has been labelled by three different students. This way we can increase the accuracy and the meaningfulness of the labelling process by selecting tweets for which all the evaluators agree on its political nature, and we employ a majority voting approach for labelling the sentiment of the tweets. Precisely, the sentiment label is set in according to a decision rule that selects the label which has a majority, that is, more than half the votes. Note that, taking into account the Krippendorff's alpha coefficient\footnote{Krippendorff's alpha coefficient~\citep{Krippendorff2003} is a statistical measure of the agreement achieved when coding a set of units of analysis in terms of the values of a variable.}, we obtain for the sentiment label $\alpha = 0.79$. The final dataset (\muacro{TData}) is then composed by $31,000$ labelled tweets. To the best of our knowledge, it represents one of the biggest dataset containing tweets classified by experts.
\subsection{Newspaper Data}
The adoption of \muacro{TData} in the training could presents some drawbacks due to the limited period it spans. For example, some important features for a classifier could be characteristic of the retrieval period and could lost their relevance accounting for a wider period. These drawbacks result in a limited ability of generalization of the model employed to classify the political tweets. To improve its generality, we built up an additional dataset (\muacro{NData}) containing all the article titles of different Italian newspapers (\textit{Repubblica}, \textit{il Manifesto}, and \textit{Libero}) so that they spanned the whole spectrum of the political points of view from the Right wing to Left one. More precisely, we selected, from the feed \muacro{RSS} history, all the articles from January the $1^{st}$ $2012$ to October the $10^{th}$ $2012$ extracting the news title, and we employed the categorization proposed by the newspaper to associate a label to the title. Precisely, if a news belongs to the political category proposed by the newspaper we set the label to $1$, otherwise $-1$. The resulting \muacro{NData} is composed by $17,388$ labelled newspaper titles, $10,670$ of which political $(61\%)$.
\subsection{Data employed for our Analysis}
\muacro{TData} and \muacro{NData} were preliminary to train a supervised methodology which can detect the tweets useful for our goal (to identify political disaffection). To obtain general results on the political disaffection we perform our analysis on a large sample of Italian Twitter community. To achieve this goal, we randomly extract $50,000$ Italian users, which have posted at least one Italian tweet\footnote{To identify if a tweet is written in Italian we employ the GuessLanguage library (https://code.google.com/p/guess-language/).} in a fixed temporal range (October $10^{th}$ to October $30^{th}$). Moreover, to extend this set to include also the less active users we perform a one-level snowball sampling~\citep{Atkinson2001}; precisely, we select for each user all its Italian followers, thus producing a set of $261,313$ users. Note that we have not expanded recursively this process to reduce the intrinsic bias produced by the snowball sampling process. Moreover, we take into account only the user profiles that has been created before April the $4^{th}$ (obtaining $167,557$ users), thus to prevent the problem of the continuous grow of the Italian Twitter community, that could affect the quality of our political disaffection investigation.
Finally, from each selected user we extract, for the period of interest (April the $4^{th}$ $2012$ to October the $10^{th}$ $2012$), all its tweets, thus producing our final set composed by $35,882,423$ tweets (\muacro{TCorpus}). 
\subsection{Public Opinion Surveys}
The public opinion surveys have been collected by \muacro{IPSOS} 
from April the $11^{th}$ $2012$ to October the $10^{th}$ $2012$.
The sampling procedure consisted in a survey through \muacro{CATI} (computer-assisted telephone interview) of a representative sample of the Italian electorate. More precisely, almost every week, respondents are contacted with a quota sampling on fixed parameters (age, gender, education) using the technique of random digit dialling.

We build two indicators for political disaffection. The first one (\muacro{NO\_VOTE}) is broader and measures the intentions not to vote at the next elections, a behavioural consequence of disaffection. It includes the percentage of survey respondents that declare to have very low intention to vote at the next elections\footnote{The total sample consists in $24,971$ respondents ($\sim 1040$ respondents per poll).} (see~\tabref{ipsos}). In details, we consider the people that answered $1$ ($1$ low - $10$ high) to the question ``How likely is it that you will vote at the next election?''.

The second indicator (\muacro{INEFFICACY}) is more specific. It measures the attitude of political inefficacy, that is to say the disbelief in the accountability of the political system and of all political parties (see~\tabref{ipsos}). We operationalize it using the propensity to vote (\muacro{PTV}) of respondents for a specific party\footnote{The total sample consists in $38,537$ respondents ($\sim 2267$ respondents per poll).} ($1$ low - $10$ high). Therefore, we include in this indicator the respondents that have low (equal to 1) \muacro{PTV} for all political parties included in the survey, or, alternatively, for those who have low (equal to $1$) \muacro{PTV}s for some political parties and missing answers for the other ones. The \muacro{PTV} has been coded for all major italian parties\footnote{ \muacro{PD} (Partito Democratico), \muacro{PDL} (Popolo delle Libert\`a.), Lega Nord, \muacro{IdV} (Italia dei Valori), \muacro{UDC} (Unione di Centro), \muacro{FLI} (Futuro e Libert\`a.), \muacro{SeL} (Sinistra Ecologia Libert\`a.), and \muacro{M5S} (Movimento 5 Stelle) }.

\begin{table}
\begin{center}
{\footnotesize
\centering
\begin{tabular}{|c|c|c|}
\hline 
$t_i$   & \muacro{INEFFICACY}   & \muacro{NO\_VOTE} \tabularnewline
\hline
\hline 
2012-04-11 & 14.86\%  & 13.23\%  \tabularnewline
\hline
2012-04-18 & 13.77\%  & 14.53\%  \tabularnewline
\hline
2012-05-02 & 22.20\%  & 19.05\%  \tabularnewline
\hline
2012-05-09 & -  & 13.37\%  \tabularnewline
\hline
2012-05-16 & 12.93\%  & 13.34\%  \tabularnewline
\hline
2012-05-23 & 16.31\%  & 12.47\%  \tabularnewline
\hline
2012-06-05 & 12.07\%  & 13.31\%  \tabularnewline
\hline
2012-06-06 & 11.03\%  & 13.76\%  \tabularnewline
\hline
2012-06-13 & -  & 10.99\%  \tabularnewline
\hline
2012-06-20 & 10.77\%  & 13.08\%  \tabularnewline
\hline
2012-06-26 & 6.91\%  & 9.29\%  \tabularnewline
\hline
2012-06-27 & 6.84\%  & 13.09\%  \tabularnewline
\hline
2012-07-04 & -  & 11.88\%  \tabularnewline
\hline
2012-07-11 & 7.87\%  & 10.04\%  \tabularnewline
\hline
2012-07-17 & 9.51\%  & 13.64\%  \tabularnewline
\hline
2012-07-18 & 6.00\%  & 10.03\%  \tabularnewline
\hline
2012-07-25 & -  & 13.26\%  \tabularnewline
\hline
2012-09-04 & -  & 13.53\%  \tabularnewline
\hline
2012-09-12 & 8.46\%  & 11.22\%  \tabularnewline
\hline
2012-09-19 & -  & 12.75\%  \tabularnewline
\hline
2012-09-25 & 9.44\%  & 12.04\%  \tabularnewline
\hline
2012-09-26 & 10.46\%  & 12.74\%  \tabularnewline
\hline
2012-10-03 & -  & 12.87\%  \tabularnewline
\hline
2012-10-10 & 11.76\%  & 14.38\%  \tabularnewline
\hline

\end{tabular}

}

\caption{Public Opinion Surveys for \muacro{INEFFICACY} and \muacro{NO\_VOTE} indicators.} \tablabel{ipsos}
\end{center}
\end{table}

\section{Classification}\seclabel{classification}
Identifying political disaffection is a complex task even for human being, so, in order to create a system for the detection of this attitude in tweets, we have to define it in a formal way. The goal is to measure disaffection in conceptually similar ways to what is measured by public opinion surveys. To that end, in order to be labelled as an expression of  political disaffection a tweet has to match the following three criteria:
\begin{itemize}
\item \textbf{Political}: the tweet should regard politics.
\item \textbf{Negative}: the sentiment of the tweet should be negative.
\item \textbf{Generic}: the message have to regard politicians or parties in general. Tweets regarding only a political party or specific politician are not considered.
\end{itemize}

\begin{figure}
\centering{}\includegraphics[width=0.75\columnwidth]{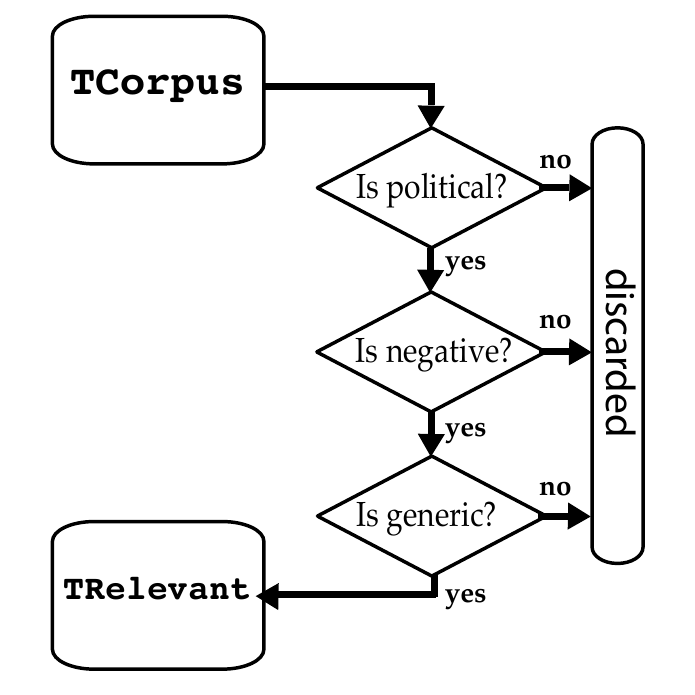}
\caption{Classification chain employed.}  \label{fig:schema_class}
\end{figure}

Since we can not train a classifier able to consider all this criteria at the same time, we created a ``chain'' of classifiers as described in Figure \ref{fig:schema_class}. There are three fundamental steps, where the relevant tweets after each step became the input for the next one. After every step the number of relevant tweets is less or equal to the number of relevant tweets after the previous step. The relevant tweets after the third step are definitively classified as relevant and all the other tweets are classified not relevant. Roughly speaking, we give \muacro{TCorpus} as input of the chain and obtain a set of tweets denoting political disaffection (\muacro{TRelevant}) as output.\\
For the first step, we trained a classification algorithm using \muacro{TData} and \muacro{NData}; the resulting classifier is able to distinguish between ``political'' and ``non-political'' tweets. In the second step, the algorithm is trained with \muacro{TData} and the resulting classifier distinguish between tweets with negative sentiment and non-negative one. The third and last step is performed by an ad-hoc classifier created with a rule-based approach to identify the generic speech.
Please note that the \muacro{TData} collection is fixed, but the features are extracted in different ways depending on classification step taken into account. 
In the next sections we summarize the feature extraction methodologies, and, subsequently, the tested classification approaches. 

\subsection{Feature extraction}
The efficacy of textual classification crucially depends on how the textual data are transformed into numerical features. 
Nevertheless, identifying the best function for feature extraction and the best tokenization method is an hard problem, and the results are usually task dependent. For these reasons, we separately manage the two supervised classification tasks: politic topics and negative sentiment. Note that, in political topics we employ both the tweets data (\muacro{TData}) and newspaper titles (\muacro{NData}).

We compared different techniques for features extraction in order to find the most suitable for our problems: 5-grams of characters, space-separated words\footnote{Considering the space-separated words' approach, we recognize as single word also emoticons, single punctuation marks such as ``!''. The URLs are also transformed in an unique token: $\langle$link$\rangle$). }, $\left\{ 1,2,3\right\}$-grams of words, and string kernels~\citep{Lodhi2002}. Concerning the counting function, we compute: term frequency~\citep{Manning2008}, boolean term presence~\citep{Mejova2011}, and term frequency-inverse document frequency (\muacro{tf-idf}, \cite{Manning2008}).  

The test was performed for the ``Politics'' topic classification using a 4-fold cross-validation with an online linear classifier. Our results show that 5-grams of characters, independently of the counting scheme, constitutes our best option.

On the other hand, taking into account the negative sentiment problem and replicating the experiments with the same methodology of the previous case, we note that the feature characterized by the space-separated word tokenization achieves the overall best results, employing as counting scheme the term frequency. 
Moreover, an important improvement is given by performing a stemming process and collapsing synonyms into a single feature; to perform this task we employ a freely available Italian synonyms dictionary\footnote{http://webs.racocatala.cat/llengua/it/sinonimi.htm}.

An important step to ensure a robust classification is to remove the sentiment target in each tweet. 
To perform this task, we employed DBpedia. This database, extracted from Wikipedia, allows to perform queries and provides a simple way to capture the semantic behind words. Then, we collect a list of possible targets for the sentiment task, by obtaining via queries a full and up-to-date set of Italian political parties, politicians, and political offices. 
Combining this data with the recognition of strings starting with ``@'' (Twitter user-names) we are able to remove the sentiment target from the tweets.

\subsection{Classification algorithms}

We need classifiers able to scale on huge corpus possibly able to be updated over time. So, we especially focused our attention on online classifiers since they require only one sweep on the data, making the classification process really fast with really good performances on the accuracy side.

We ran all the experiments on an ordinary workstation: Intel(R) Core(TM) i$7$-$2600$K CPU at $3.40$GHz with $16$Gb of RAM. 

We tested four different online algorithms for classification\footnote{Most of these algorithms are well known and have a MATLAB implementation available in DOGMA \cite{dogma}.} and one batch classification algorithm:
\begin{itemize}
\item \textbf{\muacro{ALMA}} \citep{alma}: is a fast classifier which try to approximate the maximal margin hyperplane between the two classes. We set the parameter p equals to 2 and we also tested different values of $\alpha$.
\item \textbf{\muacro{OIPCAC}} \citep{Rozza2012}: is a classification method employing a modified approach to estimate the Fisher Subspace, which allows to manage classification tasks where the space dimensionality is bigger than, or comparable to, the cardinality of the training set, and to deal with unbalanced classes.
\item \textbf{\muacro{PASSIVE AGGRESSIVE}} (\muacro{PA}, \cite{ppaa}): is a Perceptron-like method. In our experiments we test only the binary classifier with different settings.
\item \textbf{\muacro{PEGASOS}} \citep{pegasos}: is a well-know online Support Vector Machine (\muacro{SVM}) solver.
\item \textbf{\muacro{RANDOM FOREST}} (\muacro{RF}, \cite{rf}): is an algorithm based on an ensemble of classification trees. Since the algorithm is widely used in machine learning challenges with good results, we will use it as yardstick in our comparison.
\end{itemize}

We compared these algorithms on the two aforementioned tasks: 
\begin{itemize}
\item \textbf{Political}: a binary classification of tweets into ``related with politics'' and ``not related with politics''
\item \textbf{Negative}: a binary classification of tweets into ``tweets with negative sentiment'' and ``without negative sentiment'' (that is, objective or with positive sentiment). 
\end{itemize}
Note that we test the online learning algorithms in a batch setting. For this particular case, we use only their one-sweep behaviour in order to speed up the classification process.\\
After an extensive tuning of the parameters, in~\tabref{10fold-results} and in~\tabref{10-fold-neg} we report for each predictor its best performances in 10-fold cross validation on the Political classification task and on the Negative sentiment identification respectively.


\begin{table*}[th]
\begin{centering}
\begin{tabular}{|c|c|c|c|}
\hline 
Interval & $\rho$ & $95\%$ C.I. & P-Value for $\rho>0$\tabularnewline
\hline
\hline 
\textbf{$\Delta_1^{14}$} & \textbf{0.7860} & \textbf{0.476-0.922} & \textbf{0.031\%}\tabularnewline
\hline 
$\Delta_7^{14}$ & 0.7749 & 0.454-0.917 & 0.042\%\tabularnewline
\hline 
$\Delta_1^{7}$ & 0.6880 & 0.310-0.878 & 0.226\%\tabularnewline
\hline
\end{tabular}
\par\end{centering}
\caption{Pearson correlation index achieved between Twitter political disaffection and \muacro{INEFFICACY} time-series.}\tablabel{correlationgood}
\end{table*}

\begin{table*}[th]
\centering{}\begin{tabular}{|c|c|c|c|}
\hline 
Interval & $\rho$ & $95\%$ C.I. & P-Value for $\rho>0$\tabularnewline
\hline
\hline 
$\Delta_1^{14}$ & \textbf{0.5920} & \textbf{0.248-0.803} & \textbf{0.231\%}\tabularnewline
\hline 
\textbf{$\Delta_7^{14}$} & 0.5579 & 0.190-0.788 & 0.567\%\tabularnewline
\hline 
$\Delta_1^{7}$ & 0.4433 & 0.049-0.718 & 3.00\%\tabularnewline
\hline
\end{tabular}
\caption{Pearson correlation index achieved between Twitter political disaffection and \muacro{NO\_VOTE} time-series.}\tablabel{correlationbad}
\end{table*}



\begin{table} 
\begin{centering}

{\scriptsize
  \begin{tabular}
  {|c|c|c|c|}
\hline 
Classifier 
	& Accuracy 
	& F-Measure 
	& Global time \tabularnewline
\hline
\hline 
\muacro{ALMA}  
	& 0.883 $\pm$ 0.014 
	& 0.886 $\pm$ 0.011 
	& 13.5 $\pm$ 1  \tabularnewline
\hline 
\textbf{\emph{\muacro{PA}}} 
	& \textbf{\emph{0.889 $\pm$ 0.012}} 
	& \textbf{\emph{0.890 $\pm$ 0.012}} 
	& \textbf{\emph{10.62 $\pm$ 0.1}} \tabularnewline
\hline 
\muacro{PEGASOS} 
	& 0.882  $\pm$  0.010 
	& 0.883 $\pm$ 0.010 
	& 1103 $\pm$ 10 \tabularnewline
\hline 
\textbf{\muacro{OIPCAC}} 
	& \textbf{0.889  $\pm$  0.001} 
	& \textbf{0.891 $\pm$ 0.010} 
	& \textbf{5911 $\pm$ 52}   \tabularnewline
\hline
\end{tabular}
}
\caption{10-fold results for political topic detection (in bold face the best results considering F-measure). In italic the classifier selected for the classification process. With "time", we mean time employed for training and classification, in seconds. We were not able to conclude all the runs with \muacro{RF} due to its time cost.}\tablabel{10fold-results}
\end{centering}
\end{table}

\begin{table}
\begin{centering}
{\scriptsize
  \begin{tabular}{|c|c|c|c|}
\hline 
Classifier 
	& Accuracy 
	& F-Measure 
	& Global time \tabularnewline
\hline
\hline 
\textbf{\emph{\muacro{ALMA}}} 
	& \textbf{\emph{0.703 $\pm$ 0.029}} 
	& \textbf{\emph{0.745 $\pm$ 0.034}} 
	& \textbf{\emph{0.82 $\pm$ 0.28}} \tabularnewline
\hline 
\muacro{PA} 
	& 0.665 $\pm$ 0.064 
	& 0.705 $\pm$ 0.124 
	& 0.91 $\pm$ $10^{-3}$  \tabularnewline
\hline 
\muacro{PEGASOS} 
	& 0.691 $\pm$ 0.033 
	& 0.732 $\pm$  0.045 
	& 76 $\pm$ 0.1 \tabularnewline
\hline 
\muacro{OIPCAC} 
	& 0.714 $\pm$ 0.026 
	& 0.751 $\pm$ 0.024 
	& 121 $\pm$ 25 \tabularnewline
\hline 
\textbf{\muacro{RF}} 
	& \textbf{0.724 $\pm$  0.026} 
	& \textbf{0.776 $\pm$ 0.027} 
	& \textbf{2173 $\pm$ 48} \tabularnewline
\hline
\end{tabular}
}

\caption{10-fold results for political topic detection (in bold face the best results considering F-measure). In italic the classifier selected for the classification process. With "time", we mean time employed for training and classification, in seconds. }\tablabel{10-fold-neg}

\par\end{centering}{\tiny \par}
\end{table}

In order to obtain a good trade off between
accuracy and running times, we adopted a combination
of \muacro{PA} and \muacro{ALMA}. In particular we used \muacro{PA} for the
political/non-political classification and \muacro{ALMA} for the sentiment
classification part. It is also possible to obtain comparable
performances with different combinations of classifiers.

It is important to underline that since the ``Generic'' task (see~\secref{classification}) is trivial, we opt to employ an approach based on a list of a few words selected by domain experts to solve it.
\\

\section{Results}\seclabel{results}


\begin{figure*}
\centering\includegraphics[width=0.7\paperwidth]{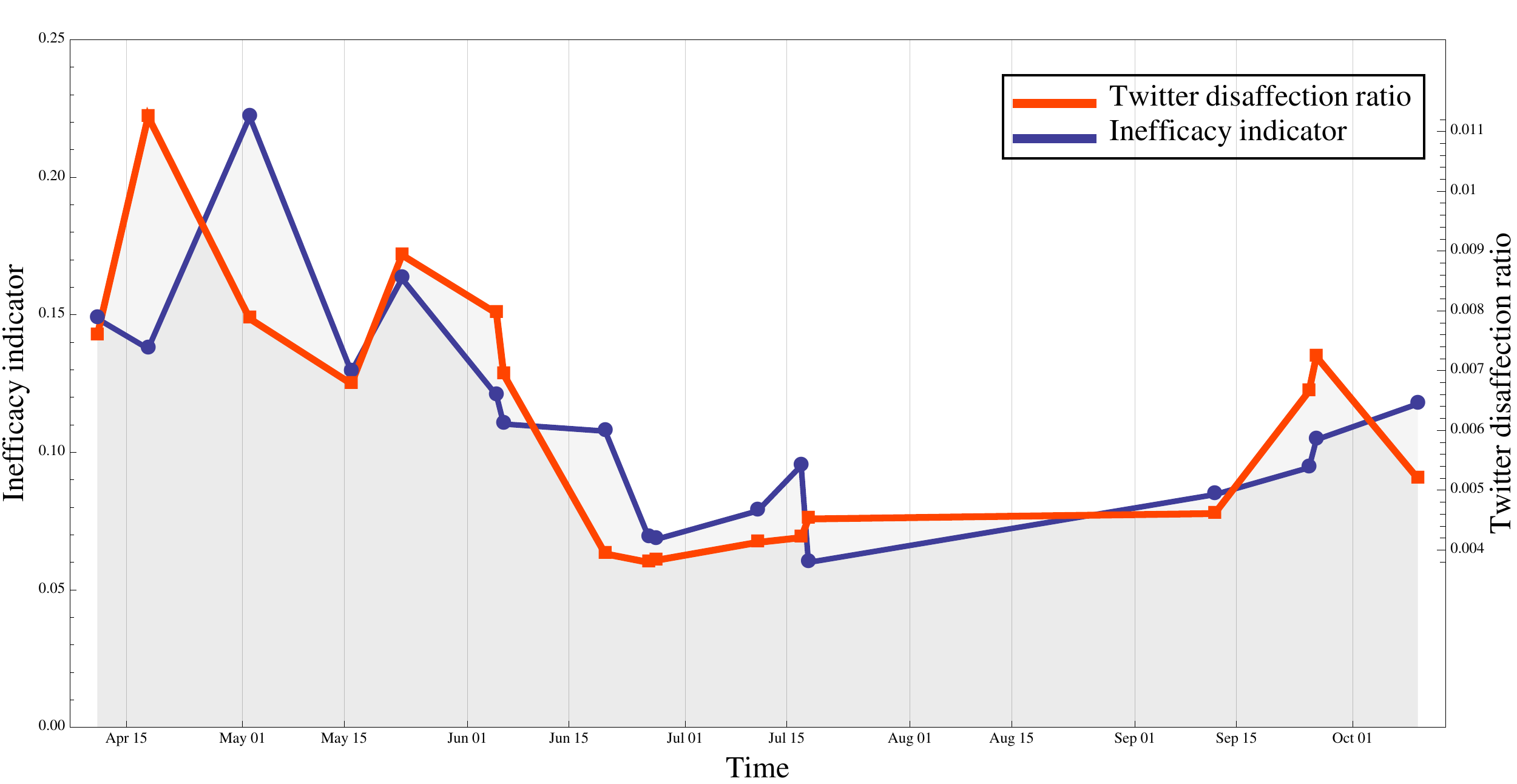}
\caption{Twitter political disaffection time-series employing $\Delta_1^{7}$ compared with the \muacro{INEFFICACY} indicator.}\figlabel{pollsvssentiment}
\end{figure*}

\begin{figure*}
\centering\includegraphics[width=0.7\paperwidth]{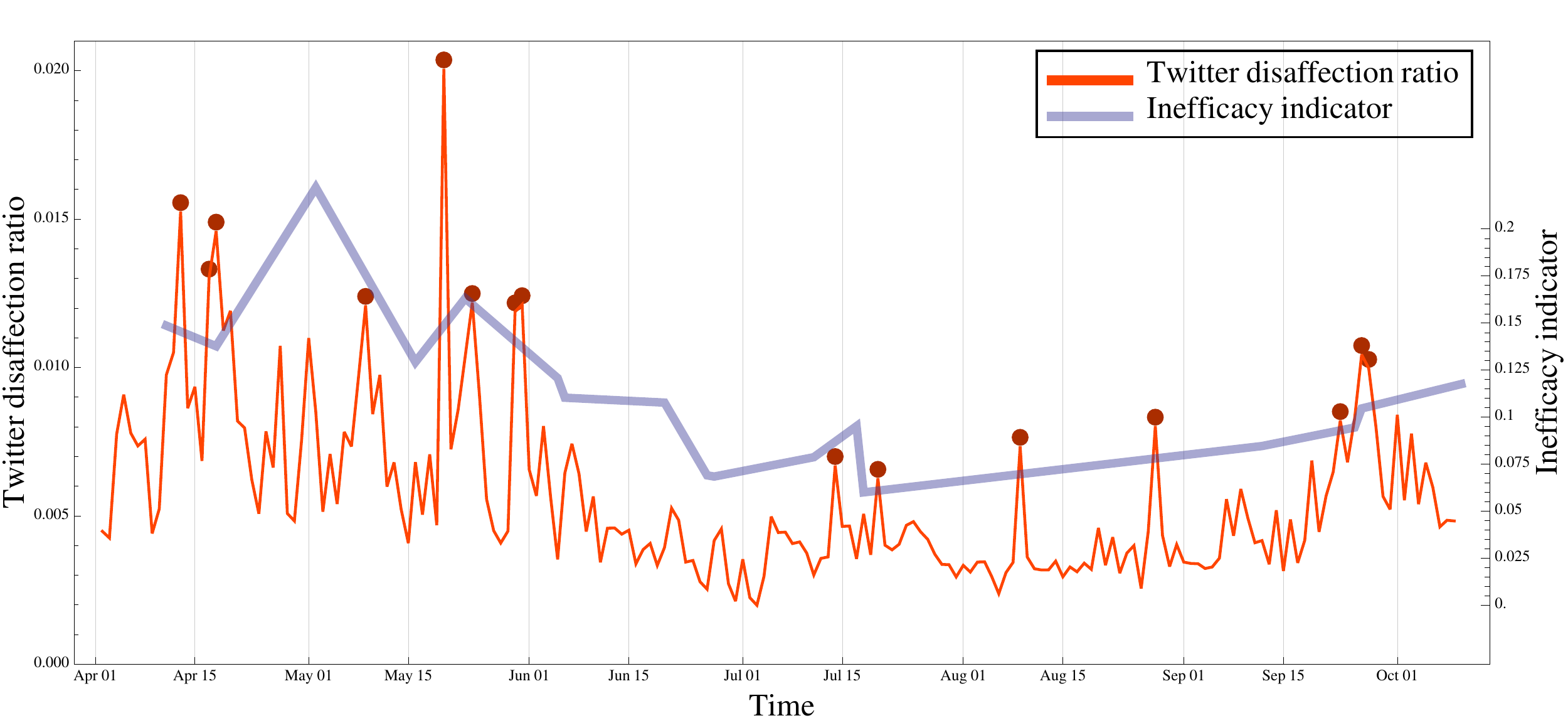}
\caption{Political disaffection tweets day by day, with the selected peaks (highlighted by circles).}\figlabel{peak}
\end{figure*}


In this section we describe the time-series obtained employing the informations extracted with the approach describes in~\secref{classification} and the relations between them and the public opinion surveys. Moreover, we summarize our methodology to identify the political news that produce the highest peaks of the generated time-series (breaking news). 

To perform a correlation analysis with the \muacro{INEFFICACY} indicator taken from surveys, we employ the approach described in~\secref{classification} to generate the set of tweets denoting political disaffection (\muacro{TRelevant}). Subsequently, taking into account each survey sampling date $t_i$ (see~\tabref{ipsos}), we generate three time-series computing the ratio between the number of political disaffection tweets and the number of political ones by employing three time intervals: 
\begin{enumerate}
\item from the date of the survey to $14$ days before ($\Delta_1^{14}$); 
\item from the day of the survey to $7$ days before ($\Delta_1^{7}$); 
\item from $7$ days before the date of the survey to $14$ before ($\Delta_7^{14}$). 
\end{enumerate}
Note that the same approach has been employed for the \muacro{NO\_VOTE} indicator (also taken from the suveys).

In~\tabref{correlationgood} it is shown the Pearson correlation index computed between the political disaffection tweet-series and the \muacro{INEFFICACY} time-series.
It is possible to underline that, the best result ($0.79$) represents a strong correlation value between \muacro{INEFFICACY} and the information extracted by our methodology. Furthermore, it is important to highlight that the best time interval is $\Delta_1^{14}$. 
This result suggests that Twitter is able to capture the change in citizens political disaffection more promptly than what public opinion surveys are able to do (as can be seen in~\figref{pollsvssentiment}). 

\begin{table}
{\centering
\tiny
\begin{tabular}{|c|p{5cm}|}
\hline
    2012-04-13 
    & \parbox{5cm}{
    	\vspace{1mm}$\checkmark$
    	La Lega prova a rifarsi un'immagine. Rinuncia agli ultimi rimborsi elettorali.\\
    	Lega tries to clear its name. It opts last electoral refunds out.

    }

\tabularnewline
\hline 
	2012-04-17
    & \parbox{5cm}{
    	\vspace{1mm}$\checkmark$
    	Lavoro, Monti pensa alla fiducia ``I partiti approveranno la riforma''.\\
    	Labor, Monti thinks of a vote of confidence ``Parties will enact reform''. 
    	
}

\tabularnewline
\hline
    2012-04-18
    & \parbox{5cm}{
    	\vspace{1mm}$\checkmark$
    	Monti: niente crescita fino al 2013, disagio lavoro per meta famiglie.\\
    	Monti: no economic growth until 2013, disadvantage for half of families.
    	
}

\tabularnewline
\hline
    2012-05-09
    & \parbox{5cm}{
    	\vspace{1mm}$\checkmark$
    	Bersani: ``Pd pi\'u forte, Monti ci ascolti'' Grillo: ``Partiti morti''. Crollo del Pdl.\\
    	Bersani: ``PD stronger, Monti listen to us'' Grillo:``Parties are dead''. PDL falls.

    	}

\tabularnewline
\hline
    2012-05-20
    & \parbox{5cm}{
    	\vspace{1mm}$\checkmark$
    	Grillo su Brindisi: strage di Stato, fa comodo a loro.\\
	Grillo about Brindisi: state massacre, it's convenient for them.
    	
	}

\tabularnewline
\hline
    2012-05-24
    & \parbox{5cm}{
    	\vspace{1mm}$\times$
    	Grillo attacca: ``Noi soldi non li vogliamo rinunceremo a rimborsi prossime politiche''.\\ 
	Grillo bashes: ``We don't need money, we opt last electoral refunds out''.

    	}

\tabularnewline
\hline
    2012-05-30
    & \parbox{5cm}{
    	\vspace{1mm}$\times$
    	Riforma Csm, il gelo di Monti cos\'i \'e fallito il piano di Catrical\'a.  \\
	CSM reform, Monti's chill, Catrical\'a's project is doomed.
    	
	}

\tabularnewline
\hline
    2012-05-31
    & \parbox{5cm}{
    	\vspace{1mm}$\times$
    	Spread, Monti resta preoccupato ``Rischio contagio malgrado gli sforzi''.\\
    	Spread, Monti worried ``Risk contagion despite the efforts''.

    	}

\tabularnewline
\hline
    2012-07-14
    & \parbox{5cm}{
    	\vspace{1mm}$\checkmark$
    	Cicchitto: ``Primarie sono inutili Berlusconi candidato premier''. \\
    	Cicchitto: ``Primary election is useless, Berlusconi is the premier candidate''.

    	}

\tabularnewline
\hline
    2012-07-19
    & \parbox{5cm}{
    	\vspace{1mm}$\times$
    	Monti ora teme il crac della Sicilia.\\
    	Now Monti is afraid of Sicily default. 

    	}

\tabularnewline
\hline
    2012-08-09
    & \parbox{5cm}{
    	\vspace{1mm}$\checkmark$
    	Monti al Wsj: ``Con Berlusconi spread a 1200'' L'ira del Pdl. E votano contro il governo. \\
	Monti to WSJ: ``Spread at 1200 with Berlusconi'' PDL anger and vote against government.
    	
}

\tabularnewline
\hline
    2012-08-28
    & \parbox{5cm}{
    	\vspace{1mm}$\checkmark$
    	Grillo a Bersani: ``Io fascista? Tu sei un fallito d'accordo con la P2''.  \\
	Grillo to Bersani: ``Am I Fascist? You're failed at one with P2''.
    	
	}

\tabularnewline
\hline
    2012-09-23
    & \parbox{5cm}{
    	\vspace{1mm}$\checkmark$
    	Cos\'i si rubava alla Regione Lazio ecco le rivelazioni di Fiorito ai pm.\\
	Fiorito's admission to public prosecutor, how they stole at Lazio government.

    	}

\tabularnewline
\hline
    2012-09-26
    & \parbox{5cm}{
    	\vspace{1mm}$\checkmark$
    	Caso Sallusti, salta l'accordo con il giudice il direttore domani rischia il carcere. \\
	Sallusti's instance, legal agreement breaks, the lead director risks the jail.
    	
	}

\tabularnewline
\hline
    2012-09-27
    & \parbox{5cm}{
    	\vspace{1mm}$\checkmark$
    	Sallusti: ``In Italia mancano le palle'' Paolo Berlusconi respinge le dimissioni.  \\
	Sallusti: ``In Italy many wimps'' Paolo Berlusconi rejects Sallusti's resignation.

    	}

\tabularnewline \hline   
\end{tabular}

\caption{Each row represents the news with the highest cosine similarity identified by our approach. The symbol $\checkmark$ represents the identified news that also appears in the political Twitter trending topic of the day taken into account. As additional contextual information, Lega is an Italian party, Bersani, Cicchitto and Berlusconi are politicians, PD is the Italian Democratic Party, Monti is the Italy premier, Sallusti and Paolo Berlusconi (Silvio Berlusconi's brother) are respectively the lead director and the editor of a newspaper, Grillo is a comic/politician, Fiorito is a regional councilman involved in bribe inquiry. CSM is the magistrates' internal board of supervisors.}\tablabel{news}
}
\end{table}

In~\tabref{correlationbad} it is shown the Pearson correlation index computed between the political disaffection tweet-series and the \muacro{NO\_VOTE} one. As can be noticed, this results ($0.59$) represents a medium correlation value but it is important because it shows that there is some connection between the modelled political disaffection and the intention to not participate at the next election day.

Taking into consideration specific affordances of twitter as a medium (i.e. immediate diffusion of news and information), these results indicate that the data that we have extracted form Twitter and data derived form surveys are two reflections of a common underlying development that exhibit different temporal characteriscs. This leads us to suggest that Twitter data could be taken as a valid alternative measurement of political disaffection.

\subsection{Breaking News Identification}
Having verified the correlation between \muacro{INEFFICACY} and the Twitter political disaffection, we have employed the \muacro{TRelevant} data to empirically determine some of the possible causes that produce the variation in disbelief in politics and politicians, hypothesising that citizens’ political inefficacy is affected by controversial political news reported daily in media.  
To achieve this goal we have identified the peaks of the time-series generated as the daily ratio between the number of political disaffection tweets and the number of political ones, and we have associated to each peak a news belonging to \muacro{NData}.

More precisely, to identify the peaks we have employed an approach similar to that proposed in~\citep{GruhlLGT04}, taking into account as peak the points of the time series greater than $\mu+2\sigma$, where $\mu$ is the mean of the points of the time-series and $\sigma$ is the standard deviation; however to improve the quality of our results we have considered for each point a set of its neighbourhood\footnote{We use a temporal window of 10 days, 5 before and 5 after the point of the time-series taken into account.} (instead of all the points) to estimate the local $\mu$s and $\sigma$s.
The qualitative results are shown in~\figref{peak}.

To associate to each peak a news, firstly we have created an inverse document frequency (\muacro{idf}) map by employing the words extracted from the corpus of the news included in \muacro{NData}, and $1,000,000$ of tweets (\muacro{PTCorpus}) randomly selected from the political subset of \muacro{TCorpus} (we employ the same classifier used for the political task as described in~\secref{classification} to identify the political tweets). Note that, these weights reduce the relevance of that terms that are recurrent in all the time-series.
For each previously identified peak we create \muacro{tf-idf} vectors for the tokenized news and tweets by employing the \muacro{idf} map, thus obtaining, for each day taken into account, two vector sets: 
\begin{itemize}
\item $\VSP{N}$ the vectors' set of the news;
\item $\VSP{T}$ the vectors' set of the tweets belonging to \muacro{PTCorpus}.
\end{itemize}
Subsequently we employed the cosine similarity between vectors to select the most correlated news w.r.t the peak taken into account as follows:
$$\argmax{\vect{n}\in \VSP{N}} \sum_{\vect{t}\in \VSP{T}}\frac{\vect{n}\cdot \vect{t}}{\left\Vert \vect{n}\right\Vert \left\Vert \vect{t}\right\Vert }$$
The results achieved are summarized in~\tabref{news}.

Finally, we have qualitatively compared the news identified with this approach with the trending topics of Twitter related to the day of each peak and we have noticed that the most of the news effectively correspond to one of the political daily trend.
However, for few peaks, \muacro{NData} does not contain any news correlated with the majority of the tweets of that day. Looking at the Twitter trending topics, we can say that this happens whenever the political discussions on Twitter do not concern about facts reported in newspapers, such as discussions spontaneously grown in the Twitter community. A meaningful example concerns the trending topic \#no2giugno: this movement asked for the suspension of the military parade of June the $2^{nd}$, seen as a waste of resources, to use these moneys to reconstruct the city of Emilia (Italian region) after the earthquakes of the $2012$. This discussion generated two peaks (May the $30^{th}$ and the $31^{st}$) that had not correlation with the traditional media news.

\section{Conclusions and Future Works}\seclabel{Conclusion}
In this work we analyse the well-known political attitude of political disaffection by using Twitter data through machine learning techniques; note that, to our knowledge, no studies have been proposed to analyze this political phenomenon employing Twitter.  
In order to validate the quality of the time-series extracted from Twitter data, we highlight the strong relation of these data with political disaffection as measured with public opinion surveys (measured through low intentions to vote and low political efficacy). 
Furthermore, we show that important political news of Italian newspapers are often correlated with the highest peaks of the produced time-series.

Note that, despite some works present many doubts about the possibility to perform an electoral prediction using Twitter data~\citep{Metaxas2011,Avello2012}, the different task of political disaffection, as suggested by the strong correlations between the time-series generated by employing the public opinion surveys and the Twitter data automatically extracted by our approach, could offer an interesting research topic for further investigations. Moreover, it is important to notice that the Twitter timeliness to answer to a political events with respect to traditional public opinion surveys suggests that our method could be employed to perform a such of daily prediction of the the citizens political disaffection change. 

In future works, we would like to extend our machinery in order to achieve better results. 
In particular we would like to integrate some interesting features from the technical point of view: the classification accuracy can be improved with a proper selection of the tweets to be labelled by experts in an active-learning fashion (i.e. \cite{bbq}); furthermore, we would like to include the information about the graph topology in order to have a better understanding of the social component of these phenomena and the possibility to employ graph-based classifiers (i.e. \cite{shazoo}, \cite{gpa}).

{
\bibliography{ICWSM13}
\bibliographystyle{aaai}
}
\end{document}